\documentclass[aps,prd,12pt,preprint,superscriptaddress,nofootinbib,floatfix]{revtex4-1}

\usepackage{latexsym}
\usepackage{amsfonts}
\usepackage[latin1]{inputenc}
\usepackage[caption=false]{subfig}
\usepackage{graphicx}
\usepackage{mathrsfs}
\usepackage{amssymb}
\usepackage{amsmath}
\usepackage{verbatim}
\usepackage{multirow}
\usepackage{color}
\usepackage{epsfig}
\usepackage{amscd}
\usepackage{bm}

\def\D0{\slash\!\!\!\!\!\!\!\!\!\:D0}

\begin{document}

\preprint{PSI-PR-13-11}
\date{\today}

\title{The Higgs sector of the 4DCHM\\ after the XLVIII Rencontres de Moriond}


\author{D.~Barducci}\email[E-mail: ]{d.barducci@soton.ac.uk}
\affiliation{School of Physics and Astronomy, University of Southampton, Southampton SO17 1BJ, U.K.}
\author{A.~Belyaev}\email[E-mail: ]{a.belyaev@soton.ac.uk}
\affiliation{School of Physics and Astronomy, University of Southampton, Southampton SO17 1BJ, U.K.}
\affiliation{Particle Physics Department, Rutherford Appleton Laboratory, Chilton, Didcot, Oxon OX11 0QX, UK}
\author{M.~S.~Brown}\email[E-mail: ]{m.s.brown@soton.ac.uk}
\affiliation{School of Physics and Astronomy, University of Southampton, Southampton SO17 1BJ, U.K.}
\author{S.~De~Curtis}\email[E-mail: ]{decurtis@fi.infn.it}
\affiliation{INFN, Sezione di Firenze, Via G. Sansone 1, 50019 Sesto Fiorentino, Italy}
\author{S.~Moretti}\email[E-mail: ]{s.moretti@soton.ac.uk}
\affiliation{School of Physics and Astronomy, University of Southampton, Southampton SO17 1BJ, U.K.}
\affiliation{Particle Physics Department, Rutherford Appleton Laboratory, Chilton, Didcot, Oxon OX11 0QX, UK}
\author{G.~M.~Pruna}\email[E-mail: ]{giovanni-marco.pruna@psi.ch}
\affiliation{Paul Scherrer Institute, CH-5232 Villigen PSI, Switzerland}

\begin{abstract}
\noindent
In this proceeding, we present the current status of a $\chi^2$ fit extracted from the profiling of the Higgs couplings performed at the LHC in the context of the 4-Dimensional Composite Higgs Model. Especially, we consider the data presented by the ATLAS and CMS collaborations during the XLVIII Rencontres de Moriond.
\end{abstract}

\maketitle

\newpage


\section{Introduction}
\label{Sec:Intro}
\noindent

Based on an alternative exploration of the Electro-Weak (EW) symmetry breaking mechanism, composite Higgs models with a Higgs state as a pseudo-Nambu-Goldstone boson (pNGB) are more and more taken into account by the physics community as one of the most elegant solutions to the Standard Model (SM) hierarchy problem (see~\cite{Kaplan:1983sm,Agashe:2004rs} for a comprehensive review). Here, we propose a framework generally known as 4-Dimensional Composite Higgs Model (4DCHM)~\cite{DeCurtis:2011yx}. Such scenario is based on the minimal coset $SO(5)/SO(4)$ which contains 4 pNGBs, one of them becoming the physical Higgs particle as a composite state from an extra-dynamic. The SM matter content is accompanied by new states: $Z$'s, $W$'s, $t$'s, $b$'s and quarks with non-standard electro-magnetic charge.

After the recent discovery by the ATLAS~\cite{Aad:2012tfa} and CMS~\cite{Chatrchyan:2012ufa} collaborations, we have devoted our efforts to the evaluation of the compatibility of such a particle with the 4DCHM. In ref.~\cite{Barducci:2013wjc,Barducci:2013wpa}, we delivered a $\chi^2$ fit involving the data of the Higgs coupling analysis at the LHC available before the XLVIII Rencontres de Moriond. Here, we provide an update of the aforementioned study in the light of the latest LHC data.


\section{Results}
\label{Sec:Results}
\noindent

To pursue our goal, we introduce the $R$ parameters:
$R_{YY}=\big[\sigma(pp\to HX)|_{\rm 4DCHM}\times {\rm BR}(H\to YY)|_{\rm 4DCHM}\big]/\big[\sigma(pp\to HX)|_{\rm SM}\times {\rm BR}(H\to YY)|_{\rm SM}\big]$,
where $YY$ identifies any possible two-body Higgs decay channel and $X$ is related to any state which is produced in association to the boson. In our previous study~\cite{Barducci:2013wjc,Barducci:2013wpa} we analysed the cases $YY=\gamma\gamma$, $b\bar b$, $WW$ and $ZZ$ in the light of the old ATLAS~\cite{ATLAS-CONF-2012-170} and CMS~\cite{CMS-PAS-HIG-12-045} data for the main Higgs production channels at the LHC: gluon-gluon fusion, Higgs-strahlung and vector-boson fusion. Then, we selected five sets of points in the 4DCHM parameter space for five different combinations of the composite scale $f$ and the coupling $g_*$ which we referred to as ``benchmarks'', each of them containing $\sim200$ points that are able to satisfy the following constraints: they must allow a reconstruction of the physical values $e$, $M_Z$, $G_F$, $m_t$, $m_b$, $m_H$, with the requirement that $e$, $M_Z$, $G_F$ lie in a range which is permitted by the Particle Data Group (PDG)~\cite{PDG}, while for the top, bottom and Higgs running masses we have used $165 ~{\rm GeV} \le m_t \le 175 ~{\rm GeV}$,  $2 ~{\rm GeV} \le m_b \le 6 ~{\rm GeV}$ and  $124 ~{\rm GeV} \le m_H \le 126 ~{\rm GeV}$. With regard to the bounds from EW precision tests, we avoided them by requiring a mass of the extra gauge bosons of 2 TeV or larger (see~\cite{Barducci:2012kk}). By making use of the latest ATLAS~\cite{ATLAS-CONF-2013-014} and CMS~\cite{CMS-PAS-HIG-13-005} data presented at the XLVIII Rencontres de Moriond, we summarise the relevant $R$ values in tab.~\ref{tab:R}.

\begin{table}[th!]
\begin{tabular}{lllll}
\hline
\ & ATLAS (old) & ATLAS (new) & ~~CMS (old) & ~~CMS (new) \\
\hline
$R_{\gamma\gamma}$ & $\phantom{-} 1.8 \pm 0.4\phantom{-}$ & $\phantom{-} 1.6 \pm 0.3\phantom{-}$ & $\phantom{-}1.564_{-0.419}^{+0.460}\phantom{-}$ & $\phantom{-}0.77 \pm 0.27$ \\
$R_{ZZ}$ & $\phantom{-} 1.0 \pm 0.4\phantom{-}$ & $\phantom{-} 1.5 \pm 0.4\phantom{-}$ & $\phantom{-}0.807_{-0.280}^{+0.349}\phantom{-}$ & $\phantom{-}0.92 \pm 0.28$\\
$R_{WW}$ & $\phantom{-} 1.5 \pm 0.6\phantom{-}$ & $\phantom{-} 1.4 \pm 0.6\phantom{-}$ & $\phantom{-}0.699_{-0.232}^{+0.245}\phantom{-}$  & $\phantom{-}0.68\pm 0.20$\\
$R_{bb}$ & $- 0.4 \pm 1.0\phantom{-}$ & $- 0.4 \pm 1.0\phantom{-}$ & $\phantom{-}1.075_{-0.566}^{+0.593}\phantom{-}$ & $\phantom{-}1.15 \pm 0.62$\\
	\hline
\end{tabular}
\caption{LHC measurements of the $R$ parameters from the ATLAS and CMS data. \label{tab:R}}
\end{table}

\begin{figure}[th!]
\includegraphics[width=0.42\textwidth]{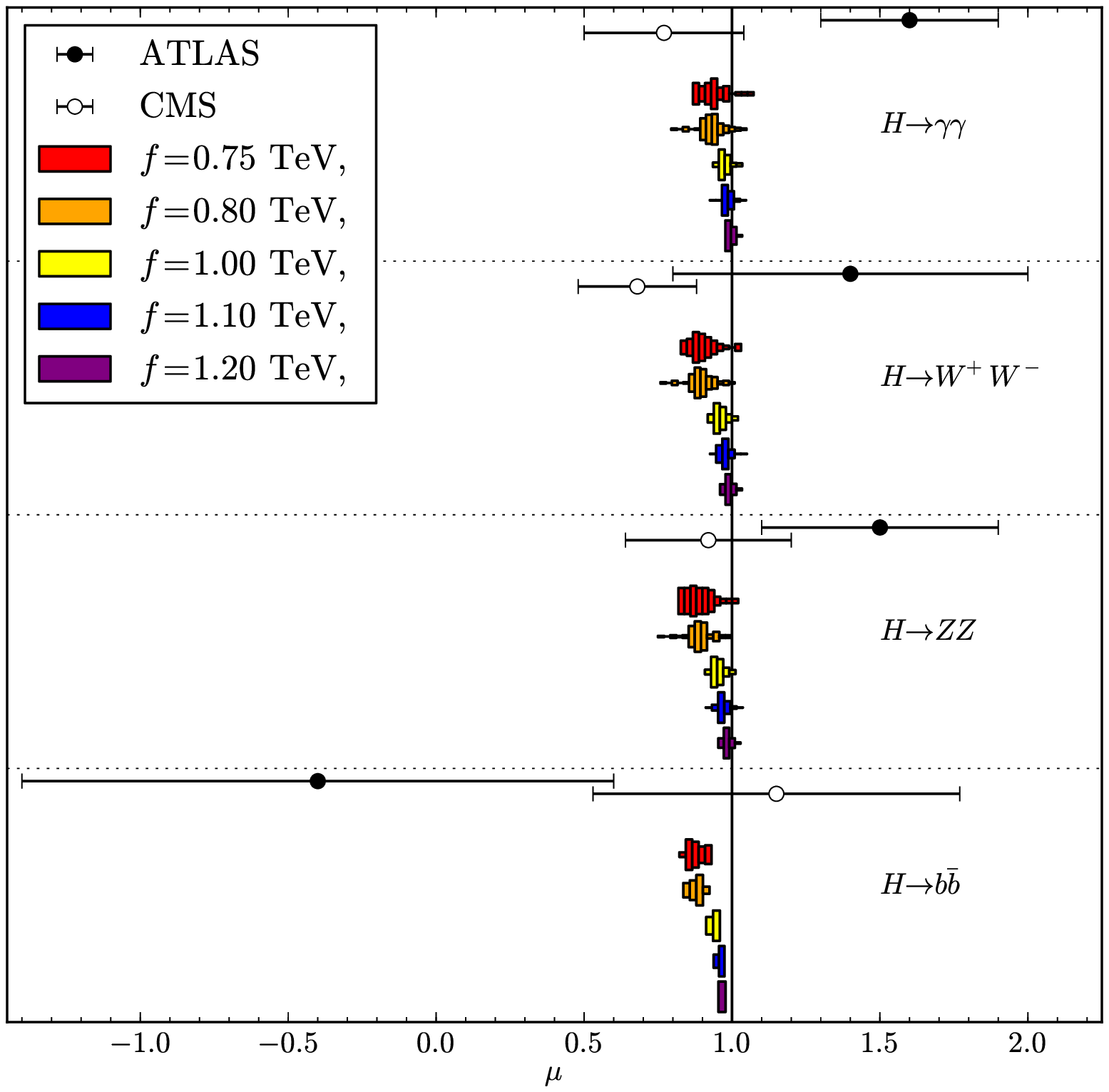}
\includegraphics[width=0.57\textwidth]{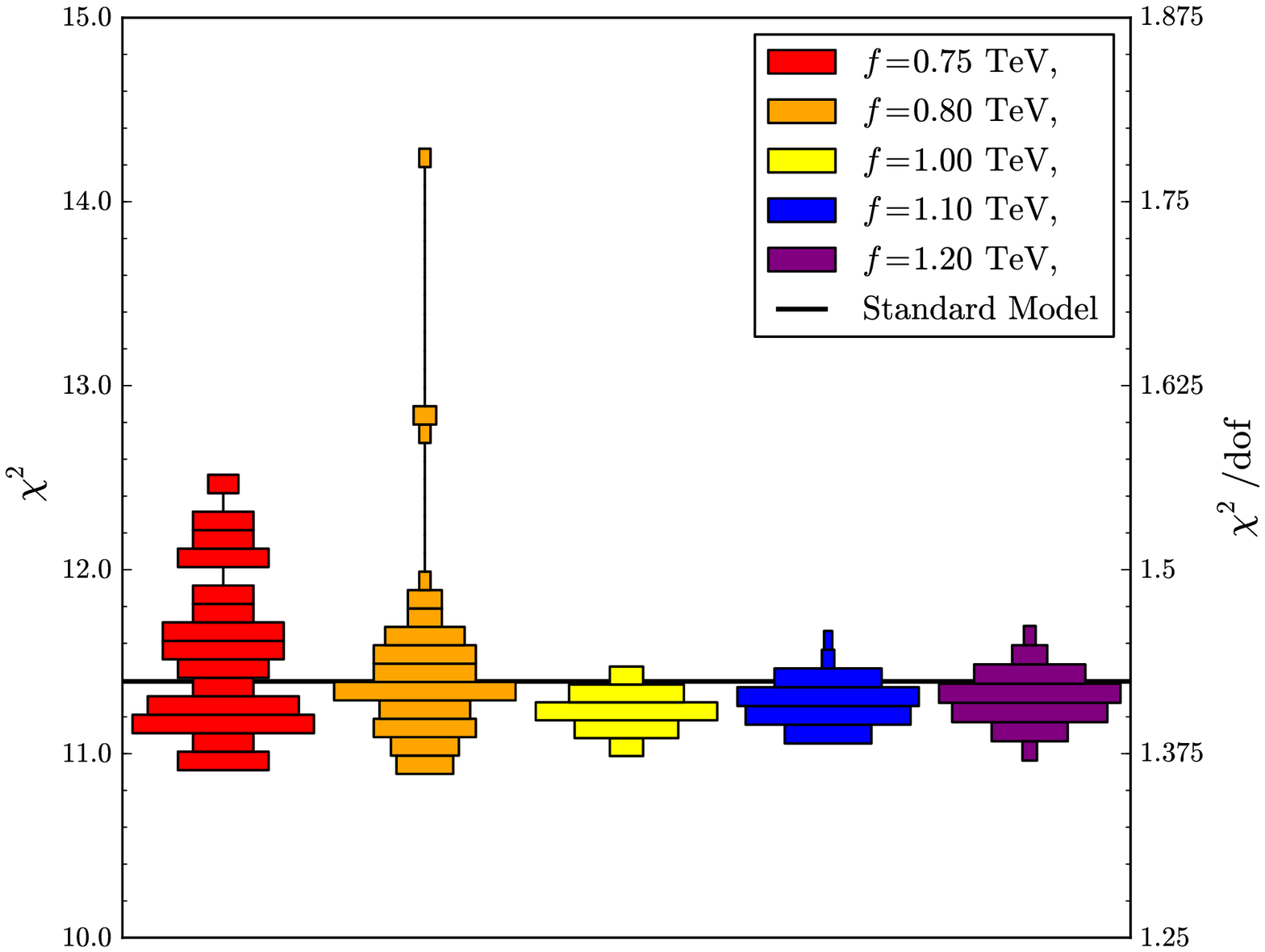}
\caption{In the left frame we plot the distribution of the benchmark sets for each Higgs boson decay channel. In the right frame we show the $\chi^2$ distribution for each benchmark.}
\label{fig:crosssections}
\end{figure}

Then, in fig.~\ref{fig:crosssections} we show the update of our previous results. In the left frame we plot the point distributions for each relevant Higgs boson decay channel, whilst in the right frame we present the $\chi^2$ distribution for each benchmark.
Carrying out a comparison among figs.~8-9 of~\cite{Barducci:2013wjc,Barducci:2013wpa} and fig.~\ref{fig:crosssections} of this note, we notice that there are no remarkable changes on the overall conclusions, nevertheless it is interesting to look closely at the impact of the new data on our fits. First of all, from tab.~\ref{tab:R} we immediately spot the differences between the old and new data: the ATLAS collaboration has lately measured a mean value of $R_{ZZ}$ which is greater than previously (well-above the SM expectation, while the old data were pointing more or less at the SM value), whereas the CMS collaboration is no longer measuring an excess in $R_{\gamma \gamma}$. In a few words, we prudently say that the status of the experimental searches reveals a tension
 between the two collaborations regarding their findings with respect to the SM expected values. As an obvious consequence, both the SM and the 4DCHM fits worsen, but such common trend operates in favour of the 4DCHM. Actually, the overall suppression of the mean values observed by CMS acts as a stabiliser for the distribution of the 4DCHM points, especially for lower values of $f$ (while higher values tend to better adhere to the SM). Still being hard to accommodate a good $\chi^2$ fit using both ATLAS and CMS data, the 4DCHM better reflects the habit for the Higgs boson in such a model to show an overall suppression of the coupling with fermions and bosons.

\section{Conclusions}
\label{Sec:Conclusions}
\noindent

In this note we updated our analysis of the Higgs couplings of the 4DCHM, using the most recent available data from ATLAS and CMS released during the XLVIII Rencontres de Moriond. Our previous conclusions (see~\cite{Barducci:2013wjc,Barducci:2013wpa}) are left unchanged.
 

\section*{Acknowledgements} 
\noindent
 
The work of GMP has been supported by the European Community's Seventh Framework Programme (FP7/2007-2013) under grant agreement n.~290605 (PSI-FELLOW/ COFUND). DB, AB and SM are financed in part through the NExT Institute.


\end{document}